\documentclass[epsfig,floats,prl,twocolumn,showpacs,preprintnumbers,floatfix]{revtex4-1}
\usepackage{graphicx}
\usepackage[latin2]{inputenc}
\usepackage{amsmath}
\usepackage{amssymb}
\usepackage{bm}
\usepackage{color}

\newcommand{\sppi}{p_{_{\text{i}}}}
\newcommand{\sppj}{p_{_{\text{j}}}}
\newcommand{\spI}{p_{_{\text{I}}}}
\newcommand{\spII}{p_{_{\text{I\!I}}}}
\newcommand{\hsi}{h_{_{\text{i}}}{\!(v)}}
\newcommand{\fti}{f\!_{_{\text{i}}}{\!(\theta)}}
\newcommand{\ftone}{f\!_{_{\text{1}}}{\!(\theta)}}
\newcommand{\ftn}{f\!_{_{\text{n}}}{\!(\theta)}}
\newcommand{\Ffti}{\int_{-{\pi}}^{{\pi}}\!\text{d}\theta\,e^{i\theta}\fti}
\newcommand{\qij}{q_{_{\text{i}\rightarrow\text{j}}}}
\newcommand{\qji}{q_{_{\text{j}\rightarrow\text{i}}}}
\newcommand{\qItoII}{q_{_{\text{I}\rightarrow\text{I\!I}}}}
\newcommand{\qIItoI}{q_{_{\text{I\!I}\rightarrow\text{I}}}}
\newcommand{\qtwoone}{q_{_{\text{2}\rightarrow\text{1}}}}
\newcommand{\qnone}{q_{_{\text{n}\rightarrow\text{1}}}}
\newcommand{\qonen}{q_{_{\text{1}\rightarrow\text{n}}}}
\newcommand{\qtwon}{q_{_{\text{2}\rightarrow\text{n}}}}
\newcommand{\qonej}{q_{_{\text{1}\rightarrow\text{j}}}}
\newcommand{\qnj}{q_{_{\text{n}\rightarrow\text{j}}}}
\newcommand{\qrt}{q_{_{\text{r}\rightarrow\text{t}}}}
\newcommand{\sqij}{\sum\limits_{j{\neq}i}\!\qij}
\newcommand{\sqonej}{\sum\limits_{j{\neq}1}\!\!\qonej}
\newcommand{\sqnj}{\sum\limits_{j{\neq}n}\!\!\qnj}
\newcommand{\Otold}{\hat{\bm u}_{t{-}\Delta t}\,{=}\!\left[\!\!\begin{array}{ccccccccc}\cos\phi \\ 
\sin\phi\end{array}\!\!\right]}
\newcommand{\Otnew}{\hat{\bm u}_t\,{=}\!\left[\!\!\begin{array}{ccccccccc}\cos\gamma \\ 
\sin\gamma\end{array}\!\!\right]}
\newcommand{\Stoldi}{S_{{_{t{-}\Delta t}},{_\text{i}}}\!(\phi)}
\newcommand{\Sti}{S_{{_t},{_\text{i}}}\!(\gamma)}
\newcommand{\Stitau}{S_{{_t},{_\text{i}}}^{\tau}\!(\gamma)}
\newcommand{\StitauF}{S_{{_t},{_\text{i}}}^{\tau}\!(m)}
\newcommand{\Stinew}{S_{{_t},{_\text{i}}}^{1}\!(\gamma)}
\newcommand{\Stitauold}{S_{{_{t{-}1}},{_\text{i}}}^{\tau{-}1}\!(\phi)}
\newcommand{\Stjtauprime}{S_{{_{t{-}1}},{_\text{j}}}^{\tau'}\!(\phi)}
\newcommand{\Snewone}{S_{{_t},{_\text{1}}}\!(\gamma)}
\newcommand{\Snewn}{S_{{_t},{_\text{n}}}\!(\gamma)}
\newcommand{\Soldone}{S_{{_{t{-}\Delta t}},{_\text{1}}}\!(\phi)}
\newcommand{\Soldn}{S_{{_{t{-}\Delta t}},{_\text{n}}}\!(\phi)}
\newcommand{\Sold}{{\bf S}_{t{-}\Delta t}}
\newcommand{\Snew}{{\bf S}_t}
\newcommand{\SnewVec}{\left(\!\!\begin{array}{ccccccccc}
\Snewone \\ \vdots \\ \Snewn \end{array}\!\!\right)}
\newcommand{\SoldVec}{\left(\!\!\begin{array}{ccccccccc}
\Soldone \\ \vdots \\ \Soldn \end{array}\!\!\right)}
\newcommand{\SoldVecFourier}{\left(\!\!\begin{array}{cc}
S_{{_{t{-}\Delta t}},{_\text{I}}}\!(m) \\
S_{{_{t{-}\Delta t}},{_\text{I\!I}}}\!(m) 
\end{array}\!\!\right)}
\newcommand{\SnewVecFourier}{\left(\!\!\begin{array}{cc}
S_{{_t},{_\text{I}}}\!(m) \\ S_{{_t},{_\text{I\!I}}}\!(m)
\end{array}\!\!\right)}
\newcommand{\MtranTT}{\!\displaystyle\int_{\!{-}\pi}^{\pi} \!\!\!\! 
d\phi \displaystyle\left[\!\!
\begin{array}{cc}
(1{-}\qItoII\!)f\!_{_{\text{I}}}{\!(\theta)}&\!\!\qIItoI\,f\!_{_{\text{I}}}{\!(\theta)}\\
\!\!\!\!\!\!\!\!\qItoII\,f\!_{_{\text{I\!I}}}{\!(\theta)}&(1{-}\qIItoI\!)f\!_{_{\text{I\!I}}}{\!(\theta)}
\end{array}
\!\!\right]\!\!.\!\!}
\newcommand{\Mtran}{\!\displaystyle\int_{\!{-}\pi}^{\pi} \!\!\!\!\!\! 
d\phi \!\!\displaystyle\left[\!\!
\begin{array}{cccc}
(1{-}\!\!\!\sqonej\!)\ftone&\!\!\qtwoone\ftone&\!\!\cdots&\!\!\qnone\ftone\\
\vdots&\!\!\ddots& &\!\!\vdots\\
\!\!\!\!\!\!\!\!\qonen\ftn&\!\!\!\qtwon\ftn&\!\!\!\!\!\cdots&\!\!\!\!\!(1{-}\!\!\!\!\sqnj\!)\ftn
\end{array}
\!\!\!\right]\!\!\!.\!\!}
\newcommand{\MtranFourier}{\displaystyle\left[\!\!
\begin{array}{cc}
(1{-}\qItoII)f\!_{_{\text{I}}}(m)&\qIItoI\,f\!_{_{\text{I}}}(m)\\
\qItoII\,f\!_{_{\text{I\!I}}}(m)&(1{-}\qIItoI)f\!_{_{\text{I\!I}}}(m)
\end{array}
\!\!\!\right]}
\newcommand{\OriCorrFunc}{\langle\hat{\bm u}_t{\cdot}\hat{\bm u}_0\rangle}
\newcommand{\tcp}{t_{\text{c}_{+}}}
\newcommand{\tcm}{t_{\text{c}_{-}}}
\newcommand{\VelCorrFunc}{\langle{\bm v}_t{\cdot}{\bm v}_0\rangle}
\newcommand{\vavg}{\langle v\rangle}

\begin{document}
\title{Orientational Memory of Active Particles in Multistate Non-Markovian 
Processes}
\author{Zeinab Sadjadi}
\author{M.\ Reza Shaebani}
\email{shaebani@lusi.uni-sb.de}
\affiliation{Department of Theoretical Physics $\&$ Center for Biophysics, 
Saarland University, D-66123 Saarbr\"ucken, Germany}

\begin{abstract}
The orientational memory of particles can serve as an effective measure of 
diffusivity, spreading, and search efficiency in complex stochastic processes. 
We develop a theoretical framework to describe the decay of directional 
correlations in a generic class of stochastic active processes consisting 
of distinct states of motion characterized by their persistence and switching 
probabilities between the states. For exponentially distributed sojourn times, 
the orientation autocorrelation is analytically derived and the characteristic 
times of its crossovers are obtained in terms of the persistence of each 
state and the switching probabilities. We show how non-exponential sojourn-time 
distributions of interest, such as Gaussian and power-law distributions, can 
result from history-dependent transitions between the states. The relaxation 
behavior of the correlation function in such non-Markovian processes is governed 
by the history-dependence of the switching probabilities and cannot be solely 
determined by the mean sojourn times of the states. 
\end{abstract}



\maketitle

Transport processes with distinct states of motion are ubiquitous in nature. 
Examples range from simple combinations of passive motility modes--- as {\it 
e.g.}\ in chromatography or transport in amorphous materials--- to more 
general mixtures of active and passive dynamics as frequently observed 
in living systems. Active dynamics of many biological agents consists of 
more than one motility mode. Migrating cells \cite{Chabaud15,dAlessandro17}, 
swimming bacteria \cite{Berg04,Najafi18}, molecular motors along cytoskeletal 
filaments \cite{Klumpp05}, and DNA-binding proteins \cite{Bauer12,Meroz09} 
are examples of agents that experience frequent transitions between two 
states of motion. While a full mathematical description of such multistate 
processes is challenging in general, a useful concept to describe the 
particle dynamics is the orientational memory, reflected {\it e.g.}\ in 
the velocity autocorrelations \cite{Peruani07}. The orientational 
correlation carries vital information about the diffusivity of the 
particle which affects its taxis, search, and transport efficiency 
\cite{Bartumeus08,Benichou11,Wadhams04}.

In order to handle multistate dynamics problems, the states are often 
approximated as simple stochastic processes, {\it e.g.}\ normal diffusion 
or ballistic motion, due to difficulties in analytical treatment of the 
full process. Although such simple mixtures have been broadly employed 
and succeeded in capturing some of the specific features of these systems 
\cite{Bressloff13,Shaebani19,Taktikos13,Jose18,Pinkoviezky13,Theves13,
Shaebani18,Watari10,Miyaguchi19,Shaebani16}, they are generally inadequate 
to accurately describe the dynamics of a combination of active states with 
arbitrary persistencies. For instance, the run-and-tumble dynamics of bacteria 
is often modeled by ballistic runs and diffusion periods or random reorientation 
events \cite{Thiel12,Angelani09,Elgeti15}. However, partial disruption of 
flagellar bundles in the tumble state results in an active swimming with 
a weak persistence rather than a pure diffusive dynamics \cite{Najafi18,
Turner16}. Moreover, ballistic motion in the run state is a very rough 
assumption since the run trajectories can be extremely curved and the 
persistence, speed, and duration of the run state vary in response to 
environmental conditions and bacterial structure \cite{Berg04,Najafi18,
Patteson15}. Thus, a full description of the bacterial dynamics requires 
a technically challenging combination of two processes with arbitrary 
self-propulsions \cite{Detcheverry17}.

The stochastic transitions between the states are often supposed to occur 
with constant probabilities, which leads to exponential sojourn-time 
distributions (as observed {\it e.g.}\ for the run-and-tumble dynamics  
of {\it E.\,coli} \cite{Taute15,Molaei14}). However, there is growing 
interest in non-exponential sojourn-time distributions. For instance, 
the run time of swarming bacteria \cite{Ariel15} or the switching time 
of the rotation direction of flagellar motors \cite{Korobkova04,Korobkova06} 
follow power-law distributions. These observations evidence age-dependent 
transitions between the states \cite{dAlessandro17,Liu18,Fedotov17}, 
which call for a detailed study of the effects of the history dependence 
of switching probabilities on sojourn-time distributions and particle 
dynamics. To design optimal navigation and taxis in non-Markovian 
active processes, a quantitative understanding of the influence of 
such memory effects on the orientational correlations is still lacking.

Here we develop a theoretical framework to quantify the orientational 
memory in multistate stochastic processes. Our approach allows us to 
calculate the orientational correlation function for arbitrary combinations 
of active and/or passive states and identify the timescales for crossovers 
of the correlation function. We verify how the exponential decay of correlations 
in processes with constant switching probabilities between the states depends 
on the persistence of the individual states and the switching probabilities. 
Moreover, we introduce specific history-dependent switching probabilities 
that lead to non-exponential sojourn-time distributions of interest, namely 
Gaussian and power-law forms. Our numerical results show that the tail behavior 
of the correlation function deviates from the exponential behavior; the 
temporal scale of the orientational memory changes in these non-Markovian 
processes with history-dependent transitions between the states. 

First, we describe the stochastic discrete process that we use to model the 
active dynamics. The process consists of $n$ distinct states of motility, 
each characterized by the probability distributions $\hsi$ and $\fti$ for 
the local speed $v$ and the directional change $\theta$ between successive 
steps of the random walk, respectively ($\text{i}{\in}\{1,{...},n\}$). We 
quantify the tendency to preserve the current direction of motion with a 
generalized self-propulsion parameter $\sppi{=}\!\Ffti$. For a turning-angle 
distribution $f(\theta)$ which is symmetric with respect to the arrival 
direction ({\it e.g.}\ walking with left-right symmetry in 2D), the 
self-propulsion reduces to $p\,{=}\,\langle\cos\theta\rangle$, {\it i.e.}\ 
a real number within $[-1,1]$. However, for the general case of an asymmetric 
$f(\theta)$, $p$ has a nonzero imaginary part as well, leading to spiral 
trajectories \cite{Sadjadi15}. $p$ is related to the anomalous exponent 
$\beta$ (describing the time evolution of the MSD) via $\beta\,{=}\,1{+}
\ln(1{+}p){/}\ln\,2$ \cite{Shaebani14}; thus, the value of $p$ reflects the 
diffusive regime of the particle dynamics: For a persistent random walk, 
$f(\theta)$ is peaked around $\theta{=}0$ ({\it i.e.}\ near forward directions) 
leading to a positive $p$ ($0{<}p{<}1$) and an anomalous exponent $\beta{>}1$ 
(superdiffusive dynamics). In the extreme case of a ballistic motion, one 
obtains $p{=}1$ and $\beta{=}2$. In contrast, $f(\theta)$ in an anti-persistent 
random walk is peaked around $\theta{=}\pi$ ({\it i.e.}\ near backward directions); 
thus, $p$ is negative (${-1}{<}p{<}0$) and $\beta{<}1$ (subdiffusive dynamics). 
A pure localization happens when the walker hopes back and forth forever, which 
leads to $p{=}{-1}$ and $\beta{=}0$. In case of normal diffusion, $f(\theta){=}
\frac{1}{2\pi}$ is isotropic which results in $p{=}0$ and $\beta{=}1$. 
Stochastic transitions between the states occur with asymmetric probabilities 
$\qij$. When a switching occurs, the walker instantly adopts the distributions 
$\hsi$ and $\fti$ of the new state. 

\begin{figure}[t]
\centerline{\includegraphics[width=0.47\textwidth]{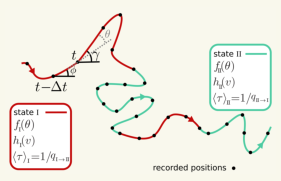}}
\caption{A sample trajectory of the persistent random walker with two states 
of motion. Each state is characterized by its turning-angle distribution 
$\fti$, speed distribution $\hsi$, and mean sojourn time $\langle \tau 
\rangle_{_{\text{i}}}$.}
\label{Fig:1}
\end{figure}

We consider an active motion in 2D in the following for brevity (extension 
to 3D is straightforward; see {\it e.g.}\ the treatment of a single-state 
persistent random walk in \cite{Sadjadi15}). As shown in the schematic 
Fig.\,\ref{Fig:1}, the orientation of the walker at successive time steps 
$t{-}\Delta t$ and $t$ is denoted with angles $\phi$ and $\gamma$, 
respectively, and the directional change $\theta$ during these two steps 
is given by $\theta{=}\gamma{-}\phi$. By introducing the orientation unit 
vectors $\Otold$ and $\Otnew$ and the probability density functions $\Stoldi$ 
and $\Sti$ to find the walker in state $\text{i}$ with the given orientation, 
the orientational state of the system at successive time steps $t{-}\Delta t$ 
and $t$ can be represented as $\Sold{=}\!\SoldVec$ and $\Snew{=}\!\SnewVec$. 
The following set of master equations describe the temporal evolution of the 
stochastic process 
\begin{eqnarray}
\Snew={\bf M}\;\Sold,
\label{Eq:MasterEqs}
\end{eqnarray}
with $\bf M$ being the transition matrix given by
\begin{eqnarray}
{\bf M}{=}\!\!\Mtran\;\;\;\;\;\;
\label{Eq:Matrix}
\end{eqnarray}
An off-diagonal element $M_{ij}$ represents the possibility of switching from 
state $j$ to $i$ with probability $\qji$ while the diagonal element $M_{ii}$ 
takes into account the possibility of remaining in state $i$ with probability 
$1{-}\!\!\sum\limits_{j{\neq}i}\!\!\qij$. The orientational change $\theta$ 
from any arbitrary direction $\phi$ to the new direction $\gamma$ is deduced 
from the turning-angle distribution $f\!_{_{\text{i}}}{\!(\theta)}$ via the 
integral $\int_{\!{-}\pi}^{\pi}d\phi\,f\!_{_{\text{i}}}{\!(\gamma{-}\phi)}$ 
over all possibilities of $\phi$. In case of a two-state process the transition 
matrix reduces to
\begin{eqnarray}
{\bf M}{=}\!\MtranTT\;\;\;\;\;\;
\label{Eq:Matrix22}
\end{eqnarray}

{\it Constant switching probabilities.---} The transitions between the states 
with constant probabilities $\qij$ lead to exponential distributions $P\!
_{_{\text{i}}}(\tau)\,{\sim}\,\text{exp}\big[\ln(1{-}\!\!\sqij\!)\;\!\tau
\big]$ for the sojourn time $\tau$ in each state with the mean sojourn times 
$\langle \tau\rangle_{_{\text{i}}}{=}1{/}\!\sqij$. For constant switching 
probabilities, we solve Eqs.\,(\ref{Eq:MasterEqs}) in Fourier space, which 
enables us to calculate the orientational correlations. While the formalism 
is developed for multi-state processes in general, hereafter we consider 
a two-state dynamics, as the most frequent multi-state process in natural 
systems (Fig.\,\ref{Fig:1}). Using the Fourier transform $S_{{_t},
{_\text{i}}}\!(m){=}\!\int_{-{\pi}}^{{\pi}}\!\text{d}\gamma\,e^{im\gamma}
S_{{_t},{_\text{i}}}\!(\gamma)$, the master equations\,(\ref{Eq:MasterEqs}) 
lead to
\begin{align}\nonumber
\!\!\!\Snew(m)&={\bf \widetilde{M}}\;\Sold(m)\\
&\equiv\!\MtranFourier\!\Sold(m),
\label{Eq:MasterEqsFourier}
\end{align}
with $\Sold(m){=}\!\SoldVecFourier$, $\Snew(m){=}\!\SnewVecFourier$, and 
$f\!_{_{\text{i}}}(m)$ being the Fourier transform of $f\!_{_{\text{i}}}
(\theta)$. Equation\,(\ref{Eq:MasterEqsFourier}) can be recursively solved to 
obtain $\Snew(m)\,{=}\,{\bf \widetilde{M}}^{^t}{\bf S}_{0}(m)$. Alternatively, 
a combined Fourier-z-transform approach \cite{Sadjadi08} can be followed 
to reach the same result.

The orientational correlation function after time $t$ can be calculated as
\begin{align}\nonumber
\!\!\!\!\!\!\!\!\!\!\!\!\!\!\!\!\!\!\!\!\!\!\!\!\!\!\!\OriCorrFunc&=
\langle \cos(\gamma_{_t}{-}\gamma_{_0})\rangle\\
&=\!\int_{-{\pi}}^{{\pi}}\!\!\!\!\!\text{d}\gamma\int_{-{\pi}}^{{\pi}}
\!\!\!\!\!\text{d}\gamma'\,\cos(\gamma{-}\gamma')\,S(\gamma,t;\gamma',
0),\,\,\,\,\,\,\,\,\,\,\,
\label{Eq:VelCorrDef}
\end{align}
where $S(\gamma,t;\gamma',0)$ is the joint probability distribution of having 
the orientation $\gamma$ and $\gamma'$ at time $t$ and $0$, respectively. 
Assuming an initial state $S_{0}(\gamma')\,{=}\,\delta(\gamma'{-}
\gamma_{_0})$, the joint probability can be written as
\begin{align}
S(\gamma,t;\gamma',0)=S(\gamma,t|\gamma',0)\,\delta(\gamma'{-}\gamma_{_0}).
\label{Eq:JointProb}
\end{align}
\begin{figure}[t]
\centerline{\includegraphics[width=0.47\textwidth]{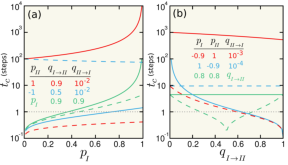}}
\caption{Characteristic time $t_\text{c}$ in terms of (a) $\spI$ and (b) 
$\qItoII$ for different values of other key parameters. The solid (dashed) 
lines represent $\tcp$ ($\tcm$) timescales. The dotted gray lines correspond 
to $t_\text{c}{=}1$, below which the particle practically carries no orientational 
memory.}
\label{Fig:2}
\end{figure}
\noindent\!\!We obtain, after some algebra, the following exact closed expression 
for the orientational correlation function
\begin{align}
\OriCorrFunc=\Big(\frac12{+}\frac{\mathcal{L}}{2\mathcal{H}}\Big)
\text{e}^{-t{/}\tcp}{+}\Big(\frac12{-}\frac{\mathcal{L}}{2\mathcal{H}}\Big)
\text{e}^{-t{/}\tcm},
\label{Eq:VelCorr}
\end{align}
with $\mathcal{L}{=}\!\!\!\!\!\!\sum\limits_{\text{i}{\in}\{\text{I},\text{I\!I}\}}
\!\!\!\!\!\!\big[(1{-}\lambda_{\text{i}}\!)(2 S\!_{_{0,\text{i}}}\!\!{-}1\!){+}
2 S\!_{_{0,\text{i}}} \qij \sppj\big]\!$, $\!\mathcal{H}{=}\sqrt{\!(\lambda_{_{
\text{I\!I}}}\!\!{-}\lambda_{_{\text{I}}}\!)^2{+}\mathcal{C}}$, $\mathcal{C}
{=}\!\!\!\!\prod\limits_{\text{i}{\in}\{\text{I},\text{I\!I}\}}\!\!\!\! 2 \qij 
\sppj$, and $\lambda_{_{\text{i}}}{=}1{-}\sppi(1{-}\qij\!)$. The initial condition 
$S\!_{_{0,\text{i}}}$--- {\it i.e.}\ the probability of initially starting in 
state $\text{i}$--- influences the orientational correlation function through 
the prefactors of the exponential terms. In the following, we choose an initially 
equilibrated system with steady probabilities $S\!_{_\text{I}}^{\;\text{st}}{=}
\frac{\qIItoI}{\qItoII{+}\qIItoI}$ and $S\!_{_\text{I\!I}}^{\;\text{st}}{=}
\frac{\qItoII}{\qItoII{+}\qIItoI}$. Note that starting from an arbitrary $S
\!_{_{0,\text{i}}}$, the Markov process of switching between the two states 
exponentially approaches the steady state with the relaxation time $t{=-1}
{/}\ln|1{-}\qItoII{-}\qIItoI|$ \cite{Hafner16}. Nevertheless, the characteristic 
times $t_{\text{c}_{\pm}}$ in Eq.\,(\ref{Eq:VelCorr}) are independent of the 
initial conditions and given by
\begin{equation}
t_{\text{c}_{\pm}}{=}{-1}\Big/\ln\Big|\frac{\mathcal{A}{\pm}\sqrt{\mathcal{A}^2
{-}4\spI\spII(1{-}\qItoII{-}\qIItoI)}}{2}\Big|,
\label{Eq:tcpm}
\end{equation}
with $\mathcal{A}\,{=}\!\!\!\!\!\sum\limits_{\text{i}{\in}\{\text{I},\text{I\!I}\}}
\!\!\!\!\!\sppi(1{-}\qij)$. The temporal scale of orientational correlations, 
set by $t_{\text{c}_{\pm}}$, can vary by several orders of magnitude by changing 
the self-propulsions $\sppi$ and switching probabilities $\qij$, as shown in 
Fig.\,\ref{Fig:2}.

\begin{figure}[t]
\centerline{\includegraphics[width=0.47\textwidth]{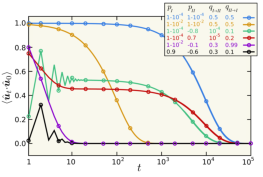}}
\caption{Orientational correlation function as a function of time for 
different values of the key parameters of the model. The symbols denote 
simulation results and the solid lines correspond to analytical predictions 
via Eq.\,(\ref{Eq:VelCorr}).}
\label{Fig:3}
\end{figure}

Equation\,(\ref{Eq:VelCorr}) implies that constant transition probabilities lead 
to an exponential decay of the orientational memory of the walker; as a result, 
the trajectory eventually gets randomized after a crossover time controlled by 
the longest characteristic time. The shape of the correlation profiles strongly 
depends on the choice of $\sppi$ and $\qij$ parameters; see Fig.\,\ref{Fig:3}. 
If the timescales $\tcp$ and $\tcm$ are well separated, the correlation function 
possesses two inflection points. For comparison of the characteristic timescales, 
$\mathcal{O}(\tcp{/}\tcm){\sim}10\;(10^3)$ for the purple (red) curve in 
Fig.\,\ref{Fig:3}. An oscillatory dynamics emerges when at least one of the 
states of motion is strongly sub-diffusive, {\it i.e.}\ has a large negative 
value of $p_i$. The particle dynamics in such a state is strongly antipersistent 
and the particle hopes frequently back and forth without a significant net 
motion. Since the direction of motion is nearly reversed at every timestep, 
the orientational correlation between successive steps is weak while between 
every two steps is strong. Similar oscillations can be observed for other 
transport properties of interest such as the mean square displacement 
\cite{Tierno12,Shaebani14}. In order to confirm the validity of the analytical 
predictions we perform extensive Monte Carlo simulations of the same stochastic 
process: A random walker in 2D with two different modes of self-propulsion is 
considered and the walker can spontaneously change the motility mode at each 
timestep according to given asymmetric switching probabilities. The simulation 
results presented in Fig.\,\ref{Fig:3} are averaged over an ensemble of $10^5$ 
realizations. The analytical predictions are in perfect agreement with the 
simulation results.

Equation\,(\ref{Eq:VelCorr}) reduces to $\text{e}^{-t{/}
t_\text{c}}$ with $t_\text{c}{=-}1{/}\ln p$ for a single-state active motion 
with self-propulsion $p$ \cite{Tierno16}. Another example is the exponential 
decay of correlations in a run-and-tumble process--- consisting of successive 
periods of ballistic run ($p_{_{\text{r}}}{=}1$) and pure diffusion ($p_{_{
\text{t}}}{=}0$)---, with a characteristic time which is purely governed by 
the run-to-tumble switching probability $\qrt$ as $t_\text{c}{=}\frac{-1}{\ln
(1{-}\qrt)}$. More generally, Eq.\,(\ref{Eq:VelCorr}) enables one to calculate 
the orientational correlation function for an arbitrary combination of two 
anomalous diffusive dynamics. Assuming an uncorrelated speed and directional 
persistence, the velocity autocorrelation can be deduced as $\VelCorrFunc\,{=}
\,\vavg\!^{^2}\OriCorrFunc$, where $\vavg\,{=}\,S\!_{_\text{I}}^{\;\text{st}}
\vavg_{_{\text{I}}}{+}S\!_{_\text{I\!I}}^{\;\text{st}}\vavg_{_{\text{I\!I}}}$; 
however, one should take into account persistence-speed correlations in general 
\cite{Maiuri15,Jerison20,Shaebani20,Wu14}. We also note that instead of the 
correlation timescale one can alternatively represent the formalism in terms 
of the correlation length scale, using the local persistence length $\ell_p$ 
extracted from $\cos(\theta){=}\text{e}^{v\Delta t{/}\ell_p}$ \cite{Landau58,
Doi86}.  

\begin{figure}[t]
\centerline{\includegraphics[width=0.47\textwidth]{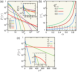}}
\caption{(a) Sojourn time distribution $P\!_{_{\text{i}}}(\tau)$ in log-log 
scale in a stochastic process with transition probabilities according to 
Eq.\,(\ref{Eq:AgeInverse}) for $q^\circ_{_{\text{i}{\rightarrow}\text{j}}}
{=}0.5$. The dashed guide line indicates a power-law decay. Inset: Same 
plot in log-lin scale. (b) Mean sojourn time in state i vs $\alpha$. 
The dashed lines denote $\langle \tau \rangle_{_{\text{i}}}$ in the 
constant-switching process with the same $q^\circ_{_{\text{i}{\rightarrow}
\text{j}}}$. (c) $P\!_{_{\text{i}}}(\tau)$ vs $\tau^2$ in a process described 
by Eq.\,(\ref{Eq:AgeExp}) for $q^\circ_{_{\text{i}{\rightarrow}\text{j}}}
{=}0$ and different values of $\tau_\text{c}$. Inset: $P\!_{_{\text{i}}}(
\tau)$ vs $\tau$ in log-lin scale.}
\label{Fig:4}
\end{figure}

{\it Age-dependent switching probabilities.---} Next we consider non-Markovian 
stochastic processes with age-dependent transition probabilities $\qij(\tau)$ 
between the states, which result in non-exponential sojourn time distributions 
$P\!_{_{\text{i}}}(\tau)$ in general. In this class of stochastic processes, 
the probability of switching from state i to j at each timestep (and thus the 
probability of remaining in state i) depends on the current residence time 
$\tau$ in state i. Therefore, we obtain the probability of a sojourn time $\tau$ 
in each state as $P\!_{_{\text{i}}}(\tau)\!=\!\frac{1}{\mathcal{N}}\!\prod
\limits_{t{=}1}^{\tau{-}1}\!\!\big(1{-}\qij(t)\big)$, with the normalization 
factor $\mathcal{N}\,{=}\,1{+}\sum\limits_{t'{=}1}^{\tau{-}1}\prod\limits_{t
{=}1}^{t'}\!\!\big(1{-}\qij(t)\big)$. Here we introduce two types of memory 
kernels which enhance or reduce the duration of stay at each state and 
modify $P\!_{_{\text{i}}}(\tau)$ towards known non-exponential forms, 
namely power-law and Gaussian distributions. The first example is an 
inverse dependence of the switching probability on the age $\tau$ of 
the state \cite{Fedotov17}. We assign a maximal probability $q^\circ_{_{
\text{i}{\rightarrow}\text{j}}}$ to switch from state i to another state 
j if the walker has just switched to state i in the previous timestep. 
However, if the walker remains in state i for a longer time, the switching 
probability to state j decreases over time according to an age-dependent 
form
\begin{align}
\qij(\tau)\,{=}\,\displaystyle\frac{q^\circ_{_{\text{i}{\rightarrow}
\text{j}}}}{\tau^\alpha}.
\label{Eq:AgeInverse}
\end{align}
We tune the history dependence via the exponent $0\,{\leq}\,\alpha\,{\leq}\,1$. 
A larger $\alpha$ leads to a faster decay of the switching probability $\qij$, 
{\it i.e.}, a longer stay in state i. While $\alpha{=}0$ corresponds to a 
constant switching probability and an exponential sojourn time distribution 
$P\!_{_{\text{i}}}(\tau)$, the limit $\alpha{=}1$ results in $P\!_{_{\text{i}}}
(\tau)\,{\propto}\,\frac{\Gamma(\tau{-}q^\circ_{_{\text{i}{\rightarrow}
\text{j}}})}{\Gamma(\tau)}$ with $\Gamma(n){=}(n{-}1)!$ being the gamma 
function. The tail of the sojourn time distribution decays as a power-law 
$P\!_{_{\text{i}}}(\tau)\,{\sim}\,\tau^{-q^\circ_{_{\text{i}{\rightarrow}
\text{j}}}}$ for which the mean sojourn time $\langle \tau \rangle_{_{\text{i}}}$ 
diverges. The gradual change of $P\!_{_{\text{i}}}(\tau)$ and $\langle \tau 
\rangle_{_{\text{i}}}$ with increasing $\alpha$ from $0$ to $1$, resulting 
from Eq.\,(\ref{Eq:AgeInverse}), is shown in Fig.\,\ref{Fig:4}(a),(b). There 
have been examples of power-law sojourn-time distributions in natural stochastic 
processes as, for example, for the run time of swarming bacteria or the switching 
time of the rotation direction of flagellar motors \cite{Ariel15,Korobkova04,
Korobkova06}. Our second choice of the history dependence is an exponentially 
saturating switching probability with the age $\tau$ of state i as
\begin{align}
\qij(\tau)\,{=}\,1-\big(1{-}q^\circ_{_{\text{i}{\rightarrow}\text{j}}}\big)\,
\text{exp}\Big[-\frac{\tau{-}1}{\tau_\text{c}}\Big].
\label{Eq:AgeExp}
\end{align}
Here, $\tau_\text{c}$ is the characteristic age and $q^\circ_{_{\text{i}{
\rightarrow}\text{j}}}$ is the minimal switching probability from state i 
to j (which applies in case of a newly started state i). The switching 
probability to state j increases with further staying in state i. In 
the limit $\tau{\rightarrow}\infty$, $\qij(\tau)$ approaches $1$, {\it 
i.e.}\ a transition from state i to j becomes highly probable. It can 
be shown that a gradual increase of the switching probability according 
to Eq.\,(\ref{Eq:AgeExp}) results in a Gaussian sojourn time distribution 
$P\!_{_{\text{i}}}(\tau)$; see Fig.\,\ref{Fig:4}(c). Note that straight 
lines in log-lin plots of $P\!_{_{\text{i}}}(\tau)$ vs $\tau^2$ represent 
a Gaussian decay. Figure\,\ref{Fig:4}(c) also shows that $P\!_{_{\text{i}}}
(\tau)$ is broader at larger values of $\tau_\text{c}$. A possible realization 
for a Gaussian sojourn-time distribution can be a tactic motion where the 
changes of the states are prevented by the level of a chemical in the 
environment which decreases exponentially over time.
  
\begin{figure}[t]
\centerline{\includegraphics[width=0.43\textwidth]{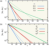}}
\caption{Comparison between the orientational correlations in processes with 
constant and history-dependent switching probabilities. The solid lines are 
$\OriCorrFunc$ for a history dependence in state\,I according to (a) 
Eq.\,(\ref{Eq:AgeInverse}) or (b) Eq.\,(\ref{Eq:AgeExp}). A constant-switching 
process is considered for state\,II. The dashed lines represent the corresponding 
results for a constant-switching process in state\,I with the same mean sojourn 
time as in the history-dependent process. $\spI\,{=}\,0.96$, $\spII\,{=}\,0.5$, 
$q^\circ_{_{\text{I}\rightarrow\text{I\!I}}}\,{=}\,\qIItoI\,{=}\,0.5$.}
\label{Fig:5}
\end{figure}

By introducing the probability density function $\Stitau$ to find the walker 
in state $\text{i}$ with age $\tau$ and orientation $\gamma$ at time $t$, 
the following master equations hold in the general case of age-dependent 
switching probabilities
\begin{equation}
\begin{cases} 
\Stitau\,{=}\,\Big(1{-}\qij(\tau{-}1)\Big)\!\!\displaystyle\int\!\!
d\phi\,f\!_{_{\text{i}}}(\gamma{-}\phi)\,\Stitauold,  
& \tau{>}1, \vspace{3mm}\\
\Stinew\,{=}\displaystyle\sum\limits_{\tau'{=}1}^{t{-}1}\qji(\tau')
\!\!\displaystyle\int\!\!d\phi\,f\!_{_{\text{i}}}(\gamma{-}\phi)\,
\Stjtauprime,  
& \tau{=}1.
\end{cases}
\label{Eq:MasterEqsNonMarkov}
\end{equation}
By recursively solving the Fourier transform of Eqs.\,(\ref{Eq:MasterEqsNonMarkov}) 
and combining them, we obtain
\begin{equation}
\StitauF{=}f\!_{_{\text{i}}}^{^{\,\tau}}\!\!(m)\Bigg[\!\prod\limits_{\text{k}
{=}1}^{\tau{-}1}\!\!\Big(1{-}\qij\!(\text{k})\Big)\!\Bigg]\sum\limits_{\tau'
{=}1}^{t{-}\tau}\!\qji(\tau')S_{{_{t{-}\tau}},{_\text{j}}}^{\tau'}\!(m). 
\end{equation}
To numerically obtain the orientational correlation function, we assume that the 
walker starts the motion with orientation $\gamma_{_0}$--- {\it i.e.}\ an initial 
state $S\!_{_{0}}(m){=}e^{i\,m\,\gamma_{_0}}$--- with an equal probability for 
being in each state. Using the Fourier transforms $f\!_{_{\text{I}}}\!(m)$ and 
$f\!_{_{\text{I\!I}}}\!(m)$ of the given turning-angle distributions, we perform 
Monte Carlo simulations with the desired forms for the age dependence of the 
switching probabilities between the states. By an ensemble of $10^6$ realizations 
of a chain of $t$ stochastic steps, we calculate $\StitauF$ and by the inverse 
Fourier transform extract the joint probability distribution of having the 
orientation $\gamma$ and $\gamma_{_0}$ at time $t$ and $0$, respectively. Then, 
following a similar procedure as described in Eqs.\,(\ref{Eq:VelCorrDef}) and 
(\ref{Eq:JointProb}), we numerically obtain the correlation function $\langle 
\cos(\gamma_{_t}{-}\gamma_{_0})\rangle$. 

The main characteristic of the orientational correlation function in 
processes with constant switching probabilities is the exponential 
relaxation according to Eq.\,(\ref{Eq:VelCorr}). The behavior is governed 
by the switching probabilities (equivalently the mean sojourn times) as 
well as the persistence of the states. According to Eqs.\,(\ref{Eq:VelCorr}) 
and (\ref{Eq:tcpm}), the long-term relaxation dynamics are dominated by 
the state with a higher persistence. To better understand the role of 
age dependence of the switching probabilities, we consider a two-state 
process with high and low persistencies and with constant $\qIItoI$ but 
history-dependent $\qItoII$ transitions. Upon increasing $\alpha$ towards 
a power-law sojourn-time distribution in the high-persistence state I, 
the orientational correlation function deviates from the exponential 
behavior and the tail becomes broader, as shown in Fig.\,\ref{Fig:5}(a). 
Nevertheless, the decay is still faster than a power-law even at $\alpha
\,{=}\,1$ (though it was recently reported that the relaxation behavior 
of the stochastic processes with power-law sojourn times may crossover 
at longer timescales \cite{Miyaguchi19}). The non-exponential form of 
the correlation function evidences that the behavior is not controlled 
by the mean sojourn times of the states anymore. For comparison, we also 
plot the correlation in a constant-switching process with the same mean 
sojourn time as in the history-dependent process; the differences become 
more pronounced with increasing $\alpha$; see dashed lines in Fig.\,\ref{Fig:5}(a). 
In a process with Gaussian sojourn time distribution in state I, Fig.\,\ref{Fig:5}(b) 
shows that the correlation decays faster than exponential (yet slower than 
Gaussian) for all choices of the characteristic history relaxation $\tau_\text{c}$. 
Similarly to the power-law age-dependence case, the behavior is not solely 
governed by the mean sojourn times of the states [see dashed lines in 
Fig.\,\ref{Fig:5}(b)]; however, the differences with a constant-switching 
process becomes negligible for $\tau_\text{c}{\rightarrow}\infty$. In this 
section we linked simple non-exponential forms of sojourn-time distribution 
(namely power-law and normal distributions) to non-Markovian transitions 
between the states of motion. More generally, realistic sojourn-time 
distributions in multi-state natural processes (that do not necessarily 
follow well-defined mathematical forms) can be also originated from (and 
numerically linked to) age-dependent switching probabilities between 
the states. Such transitions may lead to stronger or weaker orientational 
correlations, depending on whether the age-dependent switching role 
encourages or discourages a longer stay in the states. While the 
orientational memory trivially depends on the self-propulsion of the 
active agent, our findings verify that the switching statistics between 
the states can also dramatically influence the orientational memory.  

We analytically investigated the temporal evolution of orientational 
correlations in multistate active processes and derived an exact 
expression for the orientation autocorrelation in processes with 
constant switching probabilities between the states. Our theoretical 
approach opens up a new avenue to study a broad range of natural 
processes with distinct states of motility. For instance, the 
formalism can be generalized to consider stochastic dynamics with 
correlated persistence and speed as observed in cell migration 
\cite{Maiuri15,Jerison20,Shaebani20,Wu14}. It can be also extended 
to include the possibility of sharp directional changes during the 
switching events, motivated by the non-smooth tumble-to-run transitions 
commonly observed in bacterial dynamics \cite{Berg04,Najafi18}. 
Our results for age-dependent transitions between the states reveal 
that the orientational memory of the agent in such non-Markovian 
processes can be enhanced or suppressed depending on the introduced 
history dependence of the transitions. The approach is broadly 
applicable to other classes of non-Markovian stochastic processes 
with different functionalities for the age dependence of transition 
probabilities. The orientational memory of an active agent reflects 
its ability to efficiently explore the environment. Thus, our findings 
have far-reaching implications particularly for the design of optimal 
taxis, navigation, and search strategies in active systems.

We acknowledge support from the Deutsche Forschungsgemeinschaft 
(DFG) through the collaborative research center SFB 1027. MRS 
acknowledges support by Saarland University NanoBioMed initiative 
Grant No.\ 7410110401.

\bibliography{Refs}

\end{document}